\documentstyle{article}
\begin{document}
\title{\bf COVARIANT ALGEBRAIC CALCULATION
OF THE ONE-LOOP EFFECTIVE POTENTIAL
IN NON-ABELIAN GAUGE THEORY
AND  A NEW APPROACH TO STABILITY PROBLEM}

\author{I. G. Avramidi
\thanks{Alexander von Humboldt Fellow}
\thanks{This work was supported, in part by a Soros Humanitarian Foundation's
 Grant awarded by the American Physical Society and by an Award
  through the International Science Foundation's Emergency Grant
  competition.}
\thanks{ On leave of absence from Research Institute
 for Physics, Rostov State University,  Stachki 194, 344104 Rostov-on-Don,
 Russia.} \\
{ \it Department of Mathematics, University of Greifswald }\\
{\it Jahnstr. 15a, 17489 Greifswald, Germany} \\
E-mail: avramidi@mathematik.uni-greifswald.d400.de}

\date{\today}

\maketitle

\begin{abstract}

We use our  recently proposed algebraic approach for calculating the
heat kernel associated with the Laplace operator to calculate the one-loop
effective action in the non-Abelian gauge theory. We consider the
most general case of arbitrary space-time dimension, arbitrary
compact simple gauge group and arbitrary matter  and assume
 a covariantly constant gauge field strength of the most general
 form, having many independent color and space-time invariants
 (Savvidy  type  chromomagnetic vacuum) and covariantly constant
  scalar fields as a background. The explicit formulas for all the needed
   heat kernels  and zeta-functions  are obtained.
We propose a new method to study the vacuum stability and show
that  the background field configurations with covariantly constant
chromomagnetic fields can be stable only in the case when more than
one independent  field invariants are present and the values of these
invariants differ not greatly from each other.  The role of space-time
dimension is analyzed in this connection and it is shown that this is
 possible only in space-times with dimensions not less than five
 $d\geq 5$.
\end{abstract}


%%%%%%%%%%%%%%%%%%%%%%%%%%%%%%%%%%%%%%%%
\newpage

\section{Introduction}

It is well known at present that the seamy side of the asymptotic freedom in
non-Abelian gauge theories is the fact that the effective interaction
coupling becomes strong at long distances. This leads to the confinement of
color, viz. the color   objects (quarks and gluons) are `confined'  inside
hadrons and can not be observed in real physical states \cite{1,2}.
Although the perturbation theory proved to be a very successful tool at high
energies in asymptotically free theories, it is not valid any more at low
energies
 (or long distances). Here
an unknown  nonperturbative mechanism takes place that ensures the
confinement.  However, the property of confinement is not understood well
in field theory. It  was proven rigorously only in some  low-dimensional and
lattice models \cite{2} and remains still a hypothesis confirmed by the
experiments.

The main point of the confinement problem is the problem of investigation of
the vacuum structure of the theory. The vacuum seems to be far more
complicated than the  perturbation theory admits. For the  investigation of
the vacuum  problem  various approaches were proposed.  In the
phenomenological approach the general parameters of the vacuum are estimated:
non-zero vacuum expectation values of the local bilinear forms of the fields
operators (gluon and quark condensates) and, consequently, the negative
vacuum energy density,  spontaneous breakdown of the classical symmetries,
anomalies etc. \cite{3}. The effects of instantons are investigated in
\cite{4,5}.

Another possible approach is to put forward some explicit simple model, which
allows the analytical investigation. This was initiated first by Savvidy
\cite{6}
who showed that in Yang-Mills $SU(2)$ model the perturbative empty  vacuum is
unstable under the creation of a constant  chromomagnetic field that leads
to the negative vacuum energy. Such a calculation goes back to the Landau's
\cite{7} and Schwinger's \cite{8} investigations of the electron in a constant
magnetic field.  Thus the space filled with a constant homogeneous
chromomagnetic field can serve as a simple and visual model of the
nonperturbative vacuum.
The Savvidy's  result was significantly specified in further investigations
\cite{9}, where it was shown, in particular, that the Savvidy's magneto-vacuum
is
still unstable.  The real vacuum is likely to have a small domain structure
with random constant chromomagnetic fields.

The natural tool for investigating the vacuum structure is the effective
potential \cite{10}, i.e. the low-energy limit of the effective action
\cite{11}. It is
determined by general covariantly constant background fields. The
calculation of the effective action in gauge theories is complicated by the
gauge invariance. In defining the low-energy limit one has to  factorize out
the gauge degrees of freedom. That is why it is the {\it covariantly} constant
background that should be used in defining the effective potential
\cite{12,15}.

In his pioneering work \cite{6} Savvidy studied the simplest non-Abelian
 $SU(2)$ gauge group and the only possible covariantly constant
chromomagnetic field with only one nonvanishing color and space-time
 component.  However, when investigating more complicated groups
 the covariantly constant background may have {\it much more general
 structure} with {\it many independent color and space-time invariants}.
  The main aim of present paper is to investigate the effective action for
   this {\it general }case of covariantly constant background and to analyze
   some new opportunities to ensure the stability of the  vacuum that it
    provides.

We consider a Yang-Mills model with scalar and  spinor matter fields (QCD,
GUT models etc.) in flat $d$-dimensional space-time with Euclidean action
of the form

\begin{equation}
S = \int d\,x\left\{-{1\over 2g^2}{\rm tr}({\cal  F}^2_{\mu\nu}) +
i\bar\psi(\gamma^\mu\nabla_\mu+M(\varphi))\psi
+{1\over 2}\varphi^T(-\Box)\varphi+V(\varphi)\right\}
						\label{(1)}
\end{equation}

\noindent
where
$ {\cal  F}_{\mu\nu}=\partial_\mu{\cal  A}_\nu-\partial_\nu{\cal  A}_\mu+
[{\cal  A}_\mu,{\cal  A}_\nu]$ is the strength of the gauge fields taking
the values in the Lie algebra of an {\it arbitrary  compact simple} gauge
group, $g$ is the
interaction coupling constant, $\varphi=\{\varphi^A\}$ and $\psi=\{\psi^i\}$
are the multiplets of real, for definiteness, scalar fields and the Dirac
spinor
ones,
 which belong to some, in general, different irreducible representations of the
 gauge group, $M(\varphi)=\{M^i_{\ k}(\varphi)\}$ is a spinor mass matrix and
 $V(\varphi)$ is
 a potential for scalar fields, $\nabla_\mu=\partial_\mu+T({\cal  A}_\mu)$ is
the
  covariant derivative in the representation $T$ and $ \Box  =\nabla_\mu^2 $.

As we will carry the calculations in a {\it manifestly covariant} way we will
not need the explicit form of the covariant derivative. All the information
we will use is contained in the commutators of the covariant derivatives
\begin{eqnarray}
&&[\nabla_\mu, \nabla_\nu]{\cal  F}_{\alpha\beta} =
 [{\cal  F}_{\mu\nu},{\cal  F}_{\alpha\beta}] ,\nonumber\\
&&[\nabla_\mu\mu, \nabla_\nu]\psi = {\cal  R}_{\mu\nu} \psi  ,\quad
[\nabla_\mu, \nabla_\nu]\varphi = \tilde{\cal  R}_{\mu\nu} \varphi
\label{(1a)}
\end{eqnarray}
where
\begin{eqnarray}
&&{\cal  F}_{\mu\nu}=F^a_{\mu\nu}C_a  ,\nonumber \\
&&{\cal  R}_{\mu\nu} =T({\cal  F}_{\mu\nu}) = F^a_{\mu\nu}T_a  , \quad
\tilde{\cal  R}_{\mu\nu} = \tilde T({\cal  F}_{\mu\nu}) = F^a_{\mu\nu}\tilde
T_a
\label{(1b)}
\end{eqnarray}
Here $C_a=\{C^b_{\ ac}\}$, with $C^b_{\ ac}$ being the structure constants
of the gauge group,
are the generators  of the gauge group in  the adjoint representation and
$T_a$ and $\tilde T_a$ are the generators in the representations realized
 by spinor and scalar fields respectively.

To avoid misunderstanding let us make some remarks about our notations.
In this paper the {\it symbol \rm tr \it  means the trace only over group
indices},
 all other possible being left intact,
the {\it symbols \rm Sp \it  and \rm det \it  denote the trace and the
determinant
 only over vector indices} and the {\it symbol \rm $tr_\gamma$ \it  stands for
 the trace over spinor indices}. The {\it symbol \rm Tr \it  denotes the
functional
  trace}, viz. it means that not only the traces over all discrete indices
should
  be taken but over the continuous (space-time coordinates) as well.
  The explicit meaning of this notation depends on the structure of the
   quality to which it is applied. In any case the functional trace is the
trace
    over all present indices, including the continuous ones.

%%%%%%%%%%%%%%%%%%%%%%%%%%%%%%%%%%%%%%%%
\section{ One-loop effective potential}

{}From the technical point of view we are going to calculate the one-loop
effective
 action in the situation when both  the scalar background fields and the
 gauge background fields but not the spinor ones are present. The
  quantization of the model (\ref{(1)}) in a general covariant gauge leads to
  the one-loop Euclidean effective action \cite{11}
  \begin{equation}
\Gamma_{(1)} = \Gamma_{(1)YM} + \Gamma_{(1)mat}
						\label{ (2)}
\end{equation}
Here the contribution of the gauge fields (and ghosts) proper has the form
\begin{eqnarray}
&&\Gamma_{(1)YM} = {1\over 2}{\rm Tr}\ln\Delta(\lambda)/\mu^2 -
{\rm Tr}\ln F(\lambda)/\mu^2                                \label{(3)}\\
&&\Delta(\lambda) = \Delta + \lambda H                            \label{(4)}\\
&&\Delta^\mu_{\ \nu} =  - \Box
\delta^\mu_{\ \nu} - 2{\cal  F}^\mu_{\ \nu}		\label{(5)}  \\
&&H^{\mu}_{\ \nu} = \nabla^\mu\nabla_\nu                         \label{(6)}
\end{eqnarray}
where $\Delta^\mu_{\ \nu}(\lambda)$ is the inverse propagator of gauge
fields,
\begin{eqnarray}
&& F(\lambda) = \sqrt{(1-\lambda)}F                        \label{(7)}\\
&& F= -\Box     \label{(8)}
\end{eqnarray}
is the inverse ghost propagator and $\lambda$ is the gauge fixing
parameter.
 Although the  factor $\sqrt{(1-\lambda)}$ in Eq. (\ref{(7)}) seems to be
irrelevant, it
  ensures
 the gauge independence of the {\it regularized} effective  action on the
 mass shell
 (see the proof below).

The contribution of the matter fields has the form
\begin{eqnarray}
&&\Gamma_{(1)mat} = -{\rm Tr}\ln(\gamma^\mu\nabla_\mu+M(\phi))/\mu +
{1\over 2}{\rm Tr}\ln N/\mu^2                          \label{(9)}\\
&&N = -\Box +Q(\phi)                   \label{(10)}
\end{eqnarray}
where $\phi$ is the background scalar field and the mass matrix of the
scalar fields
$Q=\{Q^A_{\ B}\}$ is of the form
\begin{equation}
Q^A_{\ B}(\phi) = {\partial^2 \over \partial\phi^B\partial\phi_A}V(\phi)
 					\label{(11)}
\end{equation}

To calculate the effective {\it potential} in gauge theories, i.e. the
{\it low-energy limit} of the
effective action, one has to assume the background fields to be not
simply constant but, more
precisely, {\it covariantly} constant \cite{12}
  \begin{equation}
 \nabla_\mu{\cal  F}_{\alpha\beta} = 0 ,\quad
 \nabla_\mu M = 0 ,\quad
 \nabla_\mu Q = 0					\label{(12)}
 \end{equation}
that means, in particular,
\begin{equation}
[{\cal  F}_{\alpha\beta}, {\cal  F}_{\mu\nu}] = 0 ,\quad
[{\cal  R}_{\alpha\beta}, M] = 0 ,\quad
[\tilde{\cal  R}_{\alpha\beta}, Q] = 0	\label{(13)}
\end{equation}

Further, it is assumed that the mass matrix of fermions $M$ does not
contain the Dirac matrices or contains only even number of them, so
 that
$[M,\gamma_\mu]=0$.  Then it is easy to show, that the contribution
 of fermions
can be expressed in terms of the squared Dirac operator
\begin{equation}
{\rm Tr}\,\ln(\gamma^\mu\nabla_\mu+M)=  {1\over 2}{\rm Tr}\,\ln\,K
\label{(14)}
\end{equation}
where
\begin{eqnarray}
K&&= (\gamma^\mu\nabla_\mu+M)(-\gamma^\nu\nabla_\nu+M) \nonumber\\
   &&=-\Box- {1\over 2}\gamma^{\mu\nu}{\cal  R}_{\mu\nu} + M^2
\label{(15)}
\end{eqnarray}
and $\gamma_{\mu\nu}=\gamma_{[\mu}\gamma_{\nu]}$ are the
 generators of the orthogonal (in Euclidean case) group.

It is well known that the effective action does  not depend on the
gauge parameter $\lambda$ and on the parametrization of the
quantum fields
when the background fields lie on the mass shell, i.e.
satisfy the classical equations of motion
\cite{13}.
It is obvious that the covariantly constant background belongs to
the mass
shell, i.e does satisfy the classical equations of motion, if the scalar
fields satisfy additionally the condition of extremum
\begin{equation}
\partial  V(\phi)/\partial\phi=0					\label{(16)}
\end{equation}
and, therefore, the covariantly constant background ensures the
 independence of the effective potential on the gauge.

It is not difficult to show this explicitly.
Indeed, by differentiating $\Gamma_{(1)}$  with respect to
$\lambda$ and using
the Ward identities
\begin{eqnarray}
\nabla_\mu\Delta^{-1\ \mu}_{\ \ \quad \nu}(\lambda)
&&= - {1\over 1-\lambda}
\Box^{-1}\big(\nabla_\nu +
J_\mu\Delta^{-1\ \mu}_{\ \ \quad \nu}(\lambda)\big) 		\label{(17a)} \\
\Delta^{-1\ \mu}_{\ \ \quad \nu}(\lambda)\nabla^\nu
&&= - {1\over 1-\lambda}\big(\nabla^\mu +
\Delta^{-1\ \mu}_{\ \ \quad \nu}(\lambda)J^\nu\big)
\Box^{-1}	\label{(17b)}
\end{eqnarray}
where
\begin{equation}
J_\mu = \nabla_\nu{\cal  F}^\nu_{\ \mu}	\label{(18)}
\end{equation}
we get
\begin{equation}
{\partial \Gamma_{(1)} \over \partial \lambda}
= {1\over 2(1-\lambda)^2}{\rm Tr} \left\{\nabla_\mu J^\mu
{\Box}^{-2}
+ J_\mu\Delta^{-1\ \mu}_{\ \  \quad \nu} (\lambda)J^\nu
\Box^{-2}\right\} \label{(19)}
\end{equation}
 Hence it is obvious that on mass shell ($J=0$)  the effective potential
 really does  not depend on the gauge
 \begin{equation}
{\partial \Gamma_{(1)} \over \partial \lambda}\Bigg\vert_{J=0} = 0
						\label{(20)}
\end{equation}

Generally speaking, this proof needs a substantiation since it is  absolutely
formal, as the expressions (\ref{(3)}) and (\ref{(19)}) for the effective
action
contain the
ultraviolet divergences. More rigorously, one has to prove the gauge
 independence of the {\it regularized} effective action on mass shell.
 Using the $\zeta $ - function regularization \cite{14}
 one can rewrite the effective action in the form
 \begin{equation}
\Gamma_{(1)} = -{1\over 2}\zeta'_{tot}(0)
						\label{(21)}
\end{equation}
\begin{equation}
\zeta'(0)={d\over dp}\zeta(p)\Bigg\vert_{p=0}
						\label{(22)}
\end{equation}
where
 \begin{equation}
 \zeta_{tot}(p) = \zeta_{YM}(p) + \zeta_{mat}(p)
						\label{(23)}
 \end{equation}
 is the total  $\zeta$-function,
 \begin{equation}
 \zeta_{YM}(p) = \zeta_{\Delta(\lambda)}(p) - 2\zeta_{F(\lambda)}(p)
						\label{(24)}
  \end{equation}
 is the Yang-Mills $\zeta$-function and
 \begin{equation}
 \zeta_{mat}(p) =  - \zeta_{K}(p) + \zeta_{N}(p)
						\label{(25)}
 \end{equation}
 is the total matter fields one.

The $\zeta$-function of a differential operator $L$ is defined in terms of the
heat kernel as usual \cite{14}
 \begin{equation}
\zeta_L(p)=\mu^{2p}{\rm Tr}L^{-p}= {\mu^{2p}\over \Gamma (p)}
\int\limits_0^\infty d t\ t^{p-1}{\rm Tr} \exp(-t L)
						\label{(26)}
\end{equation}
  The $\zeta$-functions (\ref{(26)}) are analytic in the point $p=0$ and,
  therefore,
  the expression (\ref{(21)}) is finite and good defined. The matter fields
  $\zeta$-function
   does not depend on the gauge-fixing parameter $\lambda$ at all.
   Therefore,
   to prove the gauge independence of the effective action we have
   to check
   this for the Yang-Mills $\zeta$-function only.

  Let us calculate the $\zeta$-functions of the operator
  $\Delta(\lambda)$ (\ref{(4)})
  on mass shell ($J_\mu = \nabla_\nu{\cal  F}^\nu_{\ \mu}=0$).
  We have from Eq. (\ref{(4)})
  \begin{equation}
  \Delta(\lambda) = \Delta(1) - (1-\lambda)H        \label{(26a)}
  \end{equation}
  Then it is easy to show that
  \begin{equation}
  \Delta^\mu_{\ \lambda}(1)H^\lambda_{\ \nu} =
  - J^\mu\nabla_\nu \quad , \quad
  H^\mu_{\ \lambda}\Delta^\lambda_{\ \nu}(1) =
  - \nabla^\mu J_\nu    \label{(26b)}
  \end{equation}
  Therefore on mass shell (at $J_\mu=0$) the operators $\Delta(1)$
  and $H$ are
  orthogonal.  Therefrom we obtain the heat kernel for the operator
   (\ref{(26a)})
  \begin{equation}
  \exp(-t\Delta(\lambda)) = \exp(-t\Delta(1)) + \exp(t(1-\lambda)H) - 1
  \end{equation}
  Hence
  \begin{equation}
  \exp(-t\Delta(0)) = \exp(-t\Delta(1)) + \exp(tH) - 1
  \end{equation}
    which allows to express the general heat kernel in terms of the heat
    kernel for 'minimal' ($\lambda=0$) gauge operator $\Delta=\Delta(0)$
    (\ref{(5)})
  \begin{equation}
  \exp(-t\Delta(\lambda)) = \exp(-t\Delta) + \exp(t(1-\lambda)H) - \exp(tH)
							\label{(27)}
  \end{equation}
      At last making use of the relation
  \begin{equation}
  {\rm Tr}(H)^n = {\rm Tr}{\Box}^n
  \end{equation}
  which follows obviously from the definition of the operator $H$
  (\ref{(6)}),
  we express the trace of the gauge heat kernel  in terms of the minimal
  gauge heat kernel and the ghost one
  \begin{eqnarray}
  {\rm Tr}\exp(-t\Delta(\lambda)) &&= {\rm Tr}\exp(-t\Delta) \nonumber\\
  + &&{\rm Tr}\left\{\exp(-t(1-\lambda)F)
  - \exp(-t F)\right\} \label{(28)}
  \end{eqnarray}
Now by the definition of the $\zeta$-function (\ref{(26)}) we get from
Eqs. (\ref{(7)}) and (\ref{(28)})
the $\zeta$-functions for the gauge and ghost operators in general gauge
  \begin{eqnarray}
  \zeta_{\Delta(\lambda)}(p) &&= \zeta_{\Delta}(p)
  + ((1-\lambda)^{-p}-1)\zeta_F(p)         			   \\
  \zeta_{F(\lambda)}(p) &&= (1-\lambda)^{-p/2}\zeta_F(p)
 \end{eqnarray}
  Finally, by differentiating these $\zeta$-functions we obtain
  \begin{eqnarray}
  &&\zeta'_{\Delta(\lambda)}(0) = \zeta'_{\Delta}(0) -
\ln(1-\lambda)\zeta_F(0) 			\label{(29)}\\
&&\zeta'_{F(\lambda)}(0) = \zeta'_F(0) -
{1\over 2}\ln(1-\lambda)\zeta_F(0)   \label{(30)}
\end{eqnarray}
Therefore, the  $\lambda$-depending terms in the Yang-Mills
 $\zeta$-function
 (\ref{(24)}) and in the Yang-Mills effective action cancel exactly  and
 we find
\begin{eqnarray}
\zeta'_{YM}(0) &&= \zeta'_{\Delta}(0) - 2 \zeta'_{F}(0) \label{(31a)}\\
\Gamma_{(1)YM} &&= - {1\over 2}\zeta'_{\Delta}(0) + \zeta'_{F}(0)
\label{(31b)}
\end{eqnarray}
Thereby we have proven explicitly that the {\it regularized} effective
action (\ref{(21)}) does not depend on the gauge fixing parameter  $\lambda$.
Thus in what follows we choose the most convenient  so called minimal (or
diagonal) gauge by putting  $\lambda=0$.

%%%%%%%%%%%%%%%%%%%%%%%%%%%%%%%%%%%%%%%%

\section{ Heat kernel on covariantly constant background}

The operators $\Delta$ (\ref{(5)})  and $F$ (\ref{(8)}) are both of Laplace
type
\begin{equation}
L= -{\Box} + P  \label{(32)}
\end{equation}
where $ {\Box} =\nabla_\mu^2$
and $P$ is a potential term (a matrix valued function).
One can obtain the trace of the heat kernel for such operators in case of
covariantly constant background using a very elegant algebraic theorem
proven in \cite{15}.

Namely, it is proven in \cite{15} that  for any nilpotent Lie algebra given by
the
commutation relations
\begin{eqnarray}
&&[\nabla_\mu,\nabla_\nu]={\cal  R}_{\mu\nu} ,
\quad[\nabla_\mu,{\cal  R}_{\alpha\beta}]=0 ,
\quad[{\cal  R}_{\mu\nu},{\cal  R}_{\alpha\beta}]=0 \label{(32a)}\\
&&[\nabla_\mu,P]=0 ,\quad [{\cal  R}_{\mu\nu},P]=0
\end{eqnarray}
the `operator heat kernel' $\exp(-tL)$ for the `Laplace operator'
$ L=-{\Box} + P $, with $ {\Box}=\nabla_\mu^2$,
can be presented in the form of an averaging
over the Lie group
with a Gaussian measure as follows \cite{15}

\begin{eqnarray}
\exp(-t L) = &&(4\pi t)^{-d/2}
\det\left({t\hat{\cal  R}\over \sinh(t\hat{\cal  R})}\right)^{1/2}\exp(-tP)
\nonumber\\ && \times\int d k
\exp\left\{-{1\over 4t}k^\mu(t\hat{\cal  R} \coth(t\hat{\cal  R}))_{\mu\nu}
k^\nu +
k^\mu\nabla_\mu\right\}	\label{(33)}
\end{eqnarray}

Here  and everywhere below in expressions of a similar nature the
 {\it hat
denotes a matrix with vector indices}
$\hat{\cal  R}=\{{\cal  R}_{\mu\nu}\}$.The
 analytical functions of a matrix like $\hat{\cal  R}$ are always
  assumed to be
 defined in terms of corresponding power series.  In a general case
  when the matrix
  $\hat{\cal  R}$ also belongs to a representation $T$ of a Lie group, i.e.
  ${\cal  R}_{\mu\nu}= F^a_{\mu\nu}T_a $, the powers of this matrix
  should be
  understood as follows
\begin{eqnarray}
\hat{\cal  R}^0 &&= 1				\nonumber\\
\hat{\cal  R}^n &&= \{F^{a_1}_{\ \ \ \mu\lambda_1}
F^{a_2}_{\ \ \ \lambda_1\lambda_2}\cdots
               F^{a_n}_{\ \ \ \ \ \lambda_{n-1}\nu}
	       T_{a_1}T_{a_2}\cdots T_{a_n}\}
\end{eqnarray}
The `potential term' $P$ is, in general,  also a matrix with
representation indices
(they can
be not only the group ones but vector or spinor indices as well).

Acting now with the operator heat kernel (\ref{(33)}) on the $\delta$-function
 $\delta(x,x')$
 and taking the  coincidence limit  $x=x'$ one obtains finally \cite{15}
\begin{eqnarray}
&&{\rm Tr}\,\exp(-t L) \nonumber\\
&&=(4\pi t)^{-d/2}{\rm Tr}\,
\left\{\det\left({t\hat{\cal  R}\over \sinh(t\hat{\cal  R})}\right)^{1/2}
\exp(-tP)\right\}				\label{(34)}
\end{eqnarray}
This is the  generalization of the famous Schwinger's result for the Abelian
$U(1)$ gauge field (QED) \cite{8}.  As a matter of fact it is valid in much
more
general case of arbitrary semi-simple gauge group.

%%%%%%%%%%%%%%%%%%%%%%%%%%%%%%%%%%%%%

\section{Contribution of gauge fields}

Let us consider first the most essential contribution of gauge fields.
Substituting the
explicit forms of the minimal gauge operator (\ref{(5)}) and the ghost
 one (\ref{(8)})
we obtain from Eqs. (\ref{(34)})  and (\ref{(1a)}) for the traces of the heat
kernels
 appeared in previous
section

\begin{eqnarray}
&&{\rm Tr}\,\exp(-t\Delta)=\int dx (4\pi t)^{-d/2}{\rm tr}\left\{
\det\left({t\hat {\cal  F}\over \sinh(t\hat {\cal  F})}\right)^{1/2}{\rm Sp}
\exp(2t\hat{\cal  F})\right\}        \label{(35)}\\
&&{\rm Tr}\exp(-t F)=\int dx (4\pi t)^{-d/2}{\rm tr}
\det\left({t\hat {\cal  F}\over \sinh(t\hat {\cal  F})}\right)^{1/2}
\label{(36)}
\end{eqnarray}
where  $\hat{\cal  F}=\{{\cal  F}_{\mu\nu}\}$.

 Let us now calculate the traces in formulae (ref{(35)}) and (\ref{(36)}).
 The Eq. (ref{(13)}) means that the
  gauge fields take their values in the Cartan
subalgebra, i.e. the nontrivial nonvanishing components of the gauge field
exist only in the direction of the diagonal generators. The maximal number
of independent fields is equal to the dimension of the Cartan
subalgebra, i.e.
 the rank of the group $r$. Mention, first of all, that the generators
  of the Cartan
 subalgebra of the compact simple group in adjoint representation
  $C_a, (a=1,\cdots,r)$
 are the traceless antisymmetric (in real basis) commuting matrices \cite{16}
\begin{equation}
[C_a,C_b] = 0  , \quad a,b = 1,\cdots,r
\end{equation}
Hence they can be diagonalized (in complex basis) simultaneously
\begin{equation}
C_a = {\rm diag}(0,\cdots,0, i\alpha_a^{(1)}, -i\alpha_a^{(1)},
\cdots,i\alpha_a^{(p)}, -i\alpha_a^{(p)})
\end{equation}
where $\alpha^i$ are the positive roots of the algebra, $p=(n-r)/2$ is
the number of positive roots and $n$ is the dimension of the group.  The
number of zeros on the diagonal of the generators of the Cartan
subalgebra in
adjoint representation equals the maximum number of commuting
generators of
the group, i.e. the rank of the group $r$.

Therefore, the heat kernels (\ref{(35)}) and (\ref{(36)}) can be
rewritten in the form

\begin{eqnarray}
{\rm Tr}\exp(-t\Delta)=&&\int dx(4\pi t)^{-d/2}
\left\{r d
+2\sum_{\alpha>0}
\det\left({t \hat F{(\alpha)}\over \sin(t\hat  F{(\alpha)})}\right)^{1/2}
{\rm Sp}\cos(2t \hat F{(\alpha)})\right\}		\label{(37)}\\
{\rm Tr}\exp(-t F)=&&\int dx (4\pi t)^{-d/2}
\left\{r
+ 2\sum_{{\alpha}>0}
\det\left({t \hat F{(\alpha)}\over \sin(t \hat
F{(\alpha)})}\right)^{1/2}\right\}
						\label{(38)}
\end{eqnarray}

\noindent
where 2-forms
\begin{equation}
\hat F{(\alpha)} = \{ F_{\mu\nu}(\alpha)\} ,
\quad   F_{\mu\nu}(\alpha)= F^a_{\mu\nu}\alpha_a
						\label{(39)}
\end{equation}
are introduced (not to be confused with the ghost operator $F(\lambda)$
(\ref{(7)})
and the sums are to be taken over all positive roots.

To calculate the trace and the determinant over the vector indices we show
first that for any analytic function $f(\hat F)$ of a 2-form  $\hat
F=\{F_{\mu\nu}\}$
with $f(0)\ne 0$ it takes place \cite{12}
 \begin{eqnarray}
 &&{\rm Sp}f(\hat F) = (d-2q)f(0) +  \sum_{1\le j\le q}\left(f(iH_j)
+ f(-iH_j)\right)				\label{(40)}\\
&&\det f(\hat F) = (f(0))^{d-2q}\prod_{1\le j\le q}
f(iH_j)f(-iH_j) 		\label{(41)}
\end{eqnarray}
where $H_i$ are the invariants of the 2-form and $q\le [d/2]$ is the number
of independent invariants.

 The invariants $H_i$ are to be determined from the equations
 \begin{equation}
 \sum_{1\le i\le q} H_i^{2k} = I_k \qquad , \qquad  (k=1, 2, \cdots ,
[d/2])					\label{(42a)}
\end{equation}
 or, equivalently, from an algebraic equation of power $[d/2]$
 \begin{equation}
H^{2[d/2]} + c_1H^{2([d/2]-1)} + \cdots + c_{[d/2]-1}H^2 + c_{[d/2]} = 0
						\label{(42b)}
\end{equation}
 where
\begin{equation}
I_k = {1\over 2}(-1)^k{\rm Sp} \hat F^{2k}
						\label{(43)}
\end{equation}
are the basic invariants of the 2-form $\hat F$ and the coefficients
 $c_k$ have the form
\begin{equation}
c_k = \sum_{1\le j \le k}(-1)^j\sum_{{1\le k_1\le \cdots \le k_j\le
k-j+1\atop k_1+\cdots+k_j=k}}
 {1\over k_1\cdots k_j}I_{k_1}\cdots I_{k_j}
						\label{(44)}
\end{equation}
It is not difficult to show that  the coefficients $c_k$ vanish identically
for $k \ge [d/2]+1$. Besides, in the case  of Euclidean signature
all invariants $H_i$ can be regarded to be positive.

By making use of these formulae, one can compute the trace and the
determinant over vector
 indices in Eqs. (\ref{(37)}) and (\ref{(38)}) and obtain finally

 \begin{eqnarray}
{\rm Tr}\exp(-t\Delta)= &&\int dx (4\pi t)^{-d/2}
\nonumber\\
&&\times\left\{rd+ 2\sum_{\alpha>0}
\prod_{1\le i\le q}\left({tH_i(\alpha)\over \sinh(tH_i(\alpha))}\right)
\left(d + 4\sum_{1\le j\le q}\sinh^2(tH_j(\alpha))\right)\right\}
						\label{(45)}\\
{\rm Tr}\exp(-t F)= &&\int dx(4\pi t)^{-d/2}\left\{
r+2\sum_{{ \alpha}>0}
\prod_{1\le i\le q}\left({tH_i(\alpha)\over \sinh(tH_i(\alpha))}\right)\right\}
						\label{(46)}
\end{eqnarray}
where $H_i(\alpha)$  are the invariants of the tensor $ F_{\mu\nu}(\alpha)$
(\ref{(39)}).

Hence the total $\zeta$-function for gauge fields (\ref{(24)}) equals
\begin{eqnarray}
\zeta_{YM}(p) = &&
\int dx (4\pi)^{-d/2}{\mu^{2p}\over \Gamma(p)}\int\limits_0^\infty d t\,
t^{p-d/2-1}
\nonumber\\
&&\times\left\{r(d-2)+ 2\sum_{{ \alpha}>0}
\prod_{1\le i\le q}\left({tH_i(\alpha)\over \sinh(tH_i(\alpha))}\right)
\left(d-2 + 4\sum_{1\le j\le q}\sinh^2(tH_j(\alpha))\right)\right\}
                                                    \label{(47)}
\end{eqnarray}

Wherefrom it is immediately seen how the ghost fields effectively decrease
the number of degrees of freedom of  the gauge field $d\to (d-2)$.

Further calculations are possible only for a particular gauge group,
i.e. for a specific system of roots of the algebra.

%%%%%%%%%%%%%%%%%%%%%%%%%%%%%%%%%%%%%

\section{Contribution of matter fields}

Consider now the contribution of the matter fields to the one-loop effective
potential (\ref{(9)}). Using Eqs. (\ref{(34)})  and (\ref{(1a)}) one can write
down
 the heat kernels for the matter fields
 operators: the scalar (\ref{(10)}) and the spinor (\ref{(15)}) ones

 \begin{eqnarray}
 {\rm Tr}\exp(-t K)&&=\int dx (4\pi t)^{-d/2}{\rm tr}\left\{
\exp(-tM^2)\det\left({t\hat{\cal  R}\over \sinh(t\hat{\cal  R})}\right)^{1/2}
{\rm tr_\gamma}\exp\left({1\over
2}t \gamma^{\mu\nu}{\cal  R}_{\mu\nu}\right)\right\}      \label{(48)}\\
{\rm Tr}\exp(-tN)&&=\int dx (4\pi t)^{-d/2}{\rm tr}\left\{\exp(-t Q)
\det\left({t\hat{\tilde{\cal  R}}\over
\sinh(t\hat{\tilde{\cal  R}})}\right)^{1/2}\right\}           \label{(49)}
\end{eqnarray}

As above the background fields lie in the Cartan subalgebra, and, therefore,
the generators $T_a$  and $\tilde T_a$ can be diagonalized simultaneously
\begin{eqnarray}
&&T_a = {\rm diag} (i\nu_a^{(1)},\cdots,i\nu_a^{(D)}) , \nonumber\\
&&\tilde T_a = {\rm diag}
 (i\tilde\nu_a^{(1)},\cdots,i\tilde\nu_a^{(D)}) ,
\quad (a = 1,\cdots,r)			\label{(50)}
\end{eqnarray}
where $\nu$ and $\tilde\nu$ are the weights of the representations $T$
and $\tilde T$, some of them being, in general,  multiple or equal to zero
\cite{16}.

By isolating in mass matrices the singlet contributions
\begin{equation}
M^2 = M^2_{(0)} + M^{2}_aT_a\qquad,\qquad Q = Q_{(0)} + Q^a  \tilde T_a
	\label{(51)}
\end{equation}
and denoting
\begin{equation}
M^2{(\nu)} = M^{2}_a\nu_a \qquad,\qquad Q(\tilde\nu) = Q^a\tilde\nu_a
\label{(52)}
\end{equation}
we get from this for the heat kernels

\begin{eqnarray}
{\rm Tr}&&\exp(- t K) = \int dx (4\pi t)^{-d/2}\exp(-tM^2_{(0)})
 \nonumber\\ &&
\times\left\{R2^{[d/2]}+ \sum_\nu d_\nu
\exp(-i t M^2(\nu))
\det\left({t \hat F{(\nu)}\over \sin(t \hat F{(\nu)})}\right)^{1/2}
{\rm tr_\gamma}\exp\left({i\over
2}t \gamma^{\mu\nu} F_{\mu\nu}{(\nu)}\right)\right\}		\label{(53)}\\
{\rm Tr}&&\exp(- tN) = \int dx (4\pi t)^{-d/2}\exp(-tQ_{(0)})
\left\{\tilde R + \sum_{\tilde\nu}d_{\tilde\nu}\exp(-i t Q(\tilde\nu))
\det\left({t \hat F{(\tilde\nu)}\over
\sin(t \hat F{(\tilde\nu)})}\right)^{1/2}\right\}
\nonumber\\
  \label{(54)}
\end{eqnarray}

\noindent
where it is denoted $ \hat F{(\nu)} = \{F^a_{\mu\nu}\nu_a\}$,
$ \hat F{(\tilde\nu)} = \{F^a_{\mu\nu}\tilde\nu_a\}$
and the summation
is to be taken over all nonvanishing weights, $d_\nu$ and $d_{\tilde\nu}$
are the multiplicities
of the weights and  $R$  and $\tilde R$ are the numbers of zero weights, i.e.
the multiplicities
of the zero weights.

It is not difficult to get for  the trace over spinor indices \cite{12}
\begin{equation}
{\rm tr_\gamma}\exp\left\{{i\over 2}t\gamma^{\mu\nu}F_{\mu\nu}(\nu)\right\}
= 2^{[d/2]}\prod_{1\le i\le q}\cosh\left(t H_i(\nu)\right)
							\label{(55)}
\end{equation}

 Taking into account Eqs. (\ref{(41)}) and (\ref{(55)}) the final result
 for the  heat kernels
 of matter fields takes the form

 \begin{eqnarray}
 {\rm Tr}\exp(-t K) = &&\int dx (4\pi t)^{-d/2}\exp(-tM^2_{(0)})
\nonumber\\
&& \times 2^{[d/2]}
\left\{R + \sum_\nu d_\nu \exp(-i t M^2(\nu))
\prod_{1\le i \le q}\left(tH_i{(\nu)}
\coth\left(tH_i{(\nu)}\right)\right)\right\}		\label{(56)}\\
{\rm Tr}\exp(-tN) = &&\int dx (4\pi t)^{-d/2}\exp(-tQ_{(0)})
\nonumber\\
&&\times\left\{\tilde R
 + \sum_{\tilde\nu} d_{\tilde\nu}\exp(-i t Q(\tilde\nu))
\prod_{1\le i\le q}
\left({tH_i(\tilde\nu)\over \sinh\left(tH_i(\tilde\nu)\right)}\right)\right\}
                                          		\label{(57)}
\end{eqnarray}

Thus it is found for the total $\zeta-$function for matter fields (\ref{(25)})
\begin{eqnarray}
\zeta_{mat}(p) &&= \int dx (4\pi)^{-d/2}{\mu^{2p}\over \Gamma(p)}
\int\limits_0^\infty d t\,t^{p-d/2-1}
\nonumber\\
\times\Biggl\{&&
-2^{[d/2]}\exp(-tM^2_{(0)})
\left(R + \sum_\nu d_\nu\exp(-i t M^2(\nu))
\prod_{1\le i \le q}\left(tH_i{(\nu)}\coth\left(tH_i{(\nu)}
\right)\right)\right)
\nonumber\\
&&+\exp(-tQ_{(0)})\left(\tilde R +
\sum_{\tilde\nu} d_{\tilde\nu}\exp(-i t Q{\tilde\nu}) \prod_{1\le i\le
q}\left({tH_i{(\tilde\nu)}\over
\sinh\left(tH_i{(\tilde\nu)}\right)}\right)\right)\Biggr\}
						\label{(58)}
\end{eqnarray}

\noindent
where $H_i(\nu)$ and $H_i{(\tilde\nu)}$ are the invariants of the tensors
 $F^a_{\mu\nu}\nu_a$ and
$F^a_{\mu\nu}\tilde\nu_a$ respectively, defined from the
equations of the form (\ref{(42a)}) or (\ref{(42b)}).

After taking specific matter field representations one can obtain from here
more explicit expressions for the $\zeta$-functions and the effective
potential.

%%%%%%%%%%%%%%%%%%%%%%%%%%%%%%%%%%%%%

\section{ Asymptotic behavior of the heat kernel and the stability
of the vacuum}

Thus we have computed all the ingredients for the calculation of the
effective potential. By making use of obtained heat kernels one
 can get simply
the $\zeta$-functions as well as the effective potential. The exact
values of
the background fields can be determined then from the effective
equations,
viz. the condition of the extremum of the effective potential.

Let us first analyze shortly the stability of the  considered model of the
vacuum state, i.e. when the background gauge fields with covariantly
constant  field strength are present. As it is well known
the vacuum of a model
 is stable only in the case when the corresponding operators
 determining
 the dynamics of the quantum fields on the given
 background do not have
 negative modes. The presence of the negative modes
 leads obviously to
 the imaginary part of the effective potential and, as a
 consequence, to the
 instability of the vacuum. Let us stress here that it is the {\it extremum} of
the
 effective action that is important but not its sign. That is why the classical
 contribution,  which is, of course, always positive, can not
   improve the  situation
  when the negative modes pop up.

We will not derive here the explicit formulae but only analyze the problem of
negative modes. This does not require the knowledge of the whole spectrum
 and can be done very naturally by investigating only the
 {\it asymptotic behavior}
 of the heat kernels at the infinity  $t\to \infty$. Indeed,  one can show
easily
  that the behavior
 of the heat kernel for any second-order operator of Laplace type $L$ (32) on
 covariantly constant background at $t\to\infty$ is determined by the minimal
 eigenvalue $\lambda_{min}$
\begin{equation}
{\rm  Tr}\exp(-t L))\big\vert_{t\to \infty}
 \sim t^{-(d-2p)/2}\exp(-t\lambda_{min}), 			\label{(59)}
\end{equation}
$p$  being some integer. Thus one can calculate the minimal  eigenvalues of
the operators under consideration: gauge  $\Delta$, ghost $F$, Dirac $K$ and
scalar $N$ ones.

{}From Eqs. (\ref{(45)}) and (\ref{(46)}) we have

\begin{eqnarray}
&&{\rm Tr}\exp(-  t\Delta)\Big\vert_{t\to\infty}
\sim \int dx (4\pi)^{-d/2}
\nonumber\\
&&\times\left\{rd\,t^{-d/2} +
2^{q+1}
t^{q-d/2}\sum_{{ \alpha}>0}\prod_{1\le i\le q}H_i(\alpha)
\sum_{1\le k\le q}\exp\left\{-t\left(\sum_{1\le j\le q}
H_j(\alpha)-2H_k(\alpha)\right)\right\}\right\}			\label{(60)}\\
&&{\rm Tr}\exp(- t F)\Big\vert_{t\to\infty}
\sim \int dx (4\pi)^{-d/2}r\,t^{-d/2}			\label{(61)}
\end{eqnarray}

\noindent
The heat kernel for the scalar ghost operator (\ref{(61)}) behaves good, viz.
decreases, at $t\to\infty$.  This means that the minimal eigenvalue of the
operator $F$ is equal to zero
\begin{equation}
\lambda_{min}(F) = 0			\label{(62)}
\end{equation}

The heat kernel for vector gauge field operator behaves at $t\to\infty$, in
general, not good.  In this case the second term in Eq. (\ref{(60)}) can be
exponentially large.  It means that the  operator   $\Delta$ can have,
in general, the negative modes.  This is caused by the self-interaction of
the gauge  fields, (viz. by the extremely large value of the anomalous
moment of the gluon).  From Eq. (\ref{(60)}) one can conclude that the minimal
eigenvalue of $\Delta$ is either negative or equal to zero

\begin{eqnarray}
\lambda_{min}(\Delta) = \cases{\displaystyle
-\max_\alpha\left\{-\sum_{1\le i\le q}^{\ \quad\prime} H_i(\alpha)
+ H_{max}(\alpha)\right\}, 	& if this is negative      \label{(63)}\cr
 0,        					& if the previous value is positive \cr}
\end{eqnarray}

\noindent
(where $H_{max}(\alpha) = \max_{1\le i\le q}H_i(\alpha)$ is the largest
invariant of
 the field $ F_{\mu\nu}{(\alpha)}$ (\ref{(39)}), the prime at the sum meaning
that
 the sum does not include this maximal term).

Thus the vector operator $\Delta$ is positive definite and the heat kernel
behaves good at $t\to\infty$ and, consequently, the vacuum is stable,
only in the case  when the background satisfies the {\it condition of
stability}
\begin{equation}
\max_{1\le i\le q} H_i(\alpha) < \sum_{1\le i\le q}^{\ \quad\prime} H_i(\alpha)
\quad ,
 \quad {\rm for\  any} \ \alpha  \label{(63a)}
\end{equation}
i.e. if for any root $\alpha$ the maximal invariant
$H_{max}(\alpha)$ of the background field $ F_{\mu\nu}{(\alpha)}$ (\ref{(39)})
is smaller than the sum of all other ones.
This is possible only in the case when the number of independent
invariants is equal or greater than two $q\ge 2$, i.e. when the dimension
of the space is not less than four $d\ge 4$. This means that, for sure, the
 {\it chromomagnetic vacuum model is unstable in dimensions less than four.}
There are, of course,  in this case the unstable configurations too. Namely,
 those that do not satisfy the condition of stability (\ref{(63a)}).

In case $q=1$, i.e. when there is only one independent invariant of the
field  $F_{\mu\nu}{(\alpha)}$ (this case for the $SU(2)$ group was considered
first by Savvidy \cite{6}),
without fail the negative modes of the gauge field operator exist since
 the minimal
eigenvalue is negative
\begin{equation}
\lambda_{min}(\Delta) = - \max_\alpha H(\alpha) < 0
							\label{(64)}
\end{equation}
(in this case  $H(\alpha)=\sqrt{(F^a_{\mu\nu}\alpha_a)^2/2}$).
It is {\it this fact} that leads to the well known instability of the Savvidy
vacuum
\cite{9}.

Only starting with the number of invariants equal to two ($q=2$)  the
instability can ({\it but not must}!) disappear. In this case
\begin{equation}
\lambda_{min}(\Delta) = - \max_\alpha\left\{-H_{min}(\alpha)
+ H_{max}(\alpha)\right\}		\label{(65)}
\end{equation}
where $H_{min}(\alpha)$ and $H_{max}(\alpha)$ are the minimal and the maximal
invariants.  Hence it is seen, that if $H_{min}(\alpha) \ne H_{max}(\alpha)$,
then
again $\lambda_{min}(\Delta) < 0$. The only possibility to achieve the
absence of the negative modes in the case $q=2$ is to choose these
invariants equal to each other $H_{1}(\alpha)=H_{2}(\alpha)=H(\alpha)$.

Then the heat kernels (\ref{(45)}) and (\ref{(46)}) take the form

\begin{eqnarray}
{\rm Tr}\exp(-t\Delta)&&=\int dx (4\pi t)^{-d/2}\left\{
rd + 2\sum_{{\alpha}>0}
{t^2H^2(\alpha)\over \sinh^2(tH(\alpha))}
\left(d + 8\sinh^2(tH(\alpha))\right)\right\}			\label{(66)}\\
{\rm Tr}\exp(-t F)&&=\int dx (4\pi t)^{-d/2}\left\{r +
2\sum_{{ \alpha}>0}{t^2H^2(\alpha)\over \sinh^2(tH(\alpha))}\right\}
\label{(67)}
\end{eqnarray}

\noindent
and have both good decreasing power behavior at  $t\to\infty$
\begin{eqnarray}
&&{\rm Tr}\exp(-t\Delta)\Big\vert_{t\to\infty}
\sim \int dx (4\pi)^{-d/2}t^{2-d/2}(-2){\rm tr}({\cal  F}^2_{\mu\nu})
						\label{(68)}\\
&&{\rm Tr}\exp(-t F)\Big\vert_{t\to\infty}
\sim \int dx (4\pi)^{-d/2}r\,t^{-d/2}	\label{(69)}
\end{eqnarray}
Thus at $q=2, H_1(\alpha)=H_2(\alpha)$   the operator  $\Delta$ is positive
definite (except for the zero modes), i.e. the instability that is
characteristic
 to the Savvidy vacuum does not exist.
It seems on the face of it, that this case can be realized also in
four-dimensional
space-time $d=4$.  However, as we show below, in case of four dimensions
$d=4$ it is {\it not possible to
make the analytic continuation} of the equality of two invariants $H_1=H_2$
 to the
pseudo-Euclidean space of Lorentzian signature, limiting thereby the possible
physical applications of this result.

Consider the case of two invariants in {\it four-dimensional space} ($q=2,
d=4$)
at greater length.  In this case the invariants $H_i(\alpha)$ given by the
solutions
of the Eq.  (\ref{(42a)}) or (\ref{(42b)}) have the simple form
\begin{equation}
H_{1,2}(\alpha)=\sqrt{{1\over 2} I_1(\alpha) \pm {1\over 2}
\sqrt{2 I_2(\alpha) - I_1^2(\alpha)}}
\label{(70)}
\end{equation}
where $I_k(\alpha)$ are the  invariants defined by  Eq. (\ref{(43)}), viz.
\begin{eqnarray}
I_1(\alpha) &&= {1\over 2} F_{\mu\nu}(\alpha)F_{\mu\nu}(\alpha) \\
I_2(\alpha) &&= {1\over 2} F_{\mu\nu}(\alpha)F_{\nu\lambda}(\alpha)
F_{\lambda\rho}(\alpha)F_{\rho\mu}(\alpha)
\end{eqnarray}
The situation of two equal invariants  $H_1(\alpha)=H_2(\alpha)$ in four
dimensional
space considered above means, that a relation between the
invariants $I_k$ takes place
\begin{equation}
I_2(\alpha) = {1\over 2} I_1^2(\alpha)	\label{(71)}
\end{equation}
But this is possible only in space of Euclidean signature when for any field
$F_{\mu\nu}(\alpha)$
\begin{equation}
I_2(\alpha) < I_1^2(\alpha)	\label{(72)}
\end{equation}
When going to the pseudo-Euclidean (Lorentzian) signature the sign of
inequality here changes to the opposite one
\begin{equation}
I_2(\alpha)  > I_1^2(\alpha)	\label{(73)}
\end{equation}
that leads to the impossibility of the equality condition (\ref{(71)}) to be
satisfied in Minkowski space. This leads to the fact that in Euclidean case
 both invariants $H_1(\alpha)$ and $H_2(\alpha)$ are real, whereas
in Minkowski space one of them is necessarily imaginary.

Thus only beginning with  $d\ge 5$ there exists such a background that, on
 the one hand, ensures the operator $\Delta$ (5) to be  positive definite and,
  on the other hand, assumes the analytic continuation on the pseudo-Euclidean
  space of Lorentzian signature. This is a consequence of the general fact
that,
  when doing the analytic continuation of the results obtained in Euclidean
  signature to the Lorentzian
signature, one should put, in general,  one of the invariants of any 2-form
to be imaginary, i.e.
\begin{equation}
H_{q}(\alpha)=i E(\alpha)		\label{(74)}
\end{equation}
One may call the real invariants  the magnetic and the imaginary the electric
ones.
In other words, in Euclidean space all invariants are magnetic, while in the
Minkowski space one of them has to be electric (it can also vanish).

The presence of the electric field leads to poles in heat kernels,
indeterminacy in integrals over $t$ and, as a consequence, to imaginary part
of the effective potential, i. e. to the creation of the particles and
instability (although not so potent as in the presence of negative modes). It
is
not perfectly clear now how to do the analytical continuation of Euclidean
effective potential to the space of Lorentzian signature for obtaining physical
results in this general case.  The methods of the proof of the possibility of
 such a continuation \cite{17} are based essentially on the perturbative
 expansion, i.e. are valid,
strictly speaking, in weak background fields.

In contrast to the contribution of the gauge fields  the heat kernels for
matter
fields (\ref{(56)}) and (\ref{(57)}) have good exponentially decreasing
 behavior at $t\to \infty$
 (when some positive singlet contributions in mass matrices $M^2_{(0)}$ and
$Q_{(0)}$ are present). When the singlet contributions are zero or even
negative
 then the instability appear, that leads to a reconstruction of the vacuum and
to
 other values of the background fields ensuring the stability of the vacuum
state.

Thus  we have shown that in the Yang-Mills model under consideration the
Savvidy-like vacuum with constant chromomagnetic fields \cite{6,9}
can be stable only in the case when more than one constant  chromomagnetic
fields are present and the values of these fields differ not greatly from
each other.
 This is possible only in  space-times with dimensions not less than five
 $d\le 5$.

%%%%%%%%%%%%%%%%%%%%%%%%%%%%%%%%%%%%%

\section{Concluding remarks}

In this paper we continued the investigation of the low-energy effective action
 in quantum gravity and gauge theories initiated in \cite{12,15}.
  We considered the
  very wide and important class of non-Abelian gauge field theories and applied
 a very natural and elegant  pure algebraic method for calculation the
effective
 action on the covariantly constant background, developed in \cite{12}.
 The generalization
 of such algebraic approach to the case of the curved manifolds is under
 investigation in \cite{18}.

Let us  now underline shortly  our results. First of all, we considered the
most
general case of {\it arbitrary compact simple} gauge group and present the
 {\it manifestly covariant } calculation of the one-loop effective potential
for
the {\it most general covariantly constant background } with many
chromomagnetic
and space-time invariants. We showed first that the effective action does not
depend on the gauge fixing parameter and obtained then  {\it explicit }
formulas
for
all the heat kernels and zeta-functions of gauge, ghost, spinor and scalar
fields.

Further, we proposed a {\it new method to study the stability} of the vacuum
via
the asymptotic behavior of the heat kernels at the infinity. Using this method
we
found the minimal eigenvalues of all operators and analyzed  therefrom the
conditions
 of absence of negative modes. Moreover, we formulated {\it explicitly} the
  {\it condition
  of the stability} of the vacuum (as an inequality to be satisfied by the
   background fields
   (63a)). It is shown that it is impossible to get a stable vacuum of
   chromomagnetic type in space-times of dimensions less than five.
    Nevertheless, in four-dimensional space, but only
   with {\it Euclidean } signature, it is found  also a special stable
background
    field configuration (with two equal  field invariants). In higher
dimensions,
     $d>4$, there always  exist  stable field configurations, viz. those with
      many independent chromomagnetic invariants which do not differ
      greatly from each other. Of course, there are always the unstable
       configurations in this case too.

One should, perhaps, underline  here that we calculated in present paper
{\it not the  ultraviolet divergences} of the effective action, i.e. the
  beta-functions, anomalies etc. (a much discussed subject in the
  literature), but the {\it finite part} of it. While the beta-functions are
   determined in terms of the zeta-functions $\zeta(p)$  in the point
   $p=0$, we calculated much more general object, viz.
the zeta-functions at any point! That is why we did not  discuss the
asymptotic behavior of the effective coupling $g(\mu)$ based on
 the renormgroup analysis. In higher dimensions ($d>4$) the non-Abelian
 gauge theory has a dimensionful coupling constant and is above its
 critical dimension. Therefore, one could expect from the naive dimension
 consideration  that  the vacuum stabilizes  within one-loop level.
 The point is, however, not so easy. The non-Abelian gauge theory
  is {\it nonrenormalizable} in higher dimensions ($d>4$), therefore,
   one can not apply the renormgroup analysis in this case.  There
   is not any effective coupling like $g(\mu)$ in nonrenormalizable
   theories at all! That is why the behavior of considered model in
   the case $d>4$  that is found in this paper is not trivial.

%%%%%%%%%%%%%%%%%%%%%%%%%%%%%%%%%%%%%%%%

\section*{Acknowledgements}

I would like to thank G. A. Vilkovisky for many useful discussions and R.
Schimming
and J. Eichhorn for their hospitality at the Unifersity of Greifswald where
this work was
completed.

 %%%%%%%%%%%%%%%%%%%%%%%%%%%%%%%%%%%%%%%

\end{document}